# Dimensionality Reduced Antenna Array for Beamforming/steering

**Author**

Shiyi Xia[1+,*], Mingyang Zhao[2+], Qian Ma[3+], Xunnan Zhang[4+], Ling Yang[4], Yazhi Pi[5], Hyunchul Chung[3], Ad Reniers[1*], A.M.J. Koonen[1], and Zizheng Cao[5]

[1] Institute of Photonic Integration, Eindhoven University of Technology, NL 5600 MB Eindhoven, Netherlands

[2] International Collaborative Laboratory of 2D Materials for Optoelectronics Science & Technology of Ministry of Education, Institute of Microscale Optoelectronics, Shenzhen University, Shenzhen 518060, China.

[3] University of California, San Diego, CA, USA

[4] State Key Discipline Laboratory of Wide Band-gap Semiconductor Technology, School of Microelectronics, Xidian University, Xi'an 710071, China

[5] Peng Cheng Laboratory, Shenzhen 518055, China

（Correspond authors：s.xia@tue.nl, a.reniers@tue.nl）

**Abstract**

Beamforming makes possible a focused communication method. It is extensively employed in many disciplines involving electromagnetic waves, including arrayed ultrasonic, optical, and high-speed wireless communication. Conventional beam steering often requires the addition of separate active amplitude phase control units after each radiating element. The high power consumption and complexity of large-scale phased arrays can be overcome by reducing the number of active controllers, pushing beamforming into satellite communications and deep space exploration. Here, we suggest a brand-new design for a phased array antenna with a dimension reduced cascaded angle offset (DRCAO-PAA). Furthermore, the suggested DRCAO-PAA was compressed by using the concept of singular value deposition. To pave the way for practical application the particle swarm optimization algorithm and deep neural network Transformer were adopted. Based on this theoretical framework, an experimental board was built to verify the theory. Finally, the 16/8/4 -array beam steering was demonstrated by using 4/3/2 active controllers, respectively.

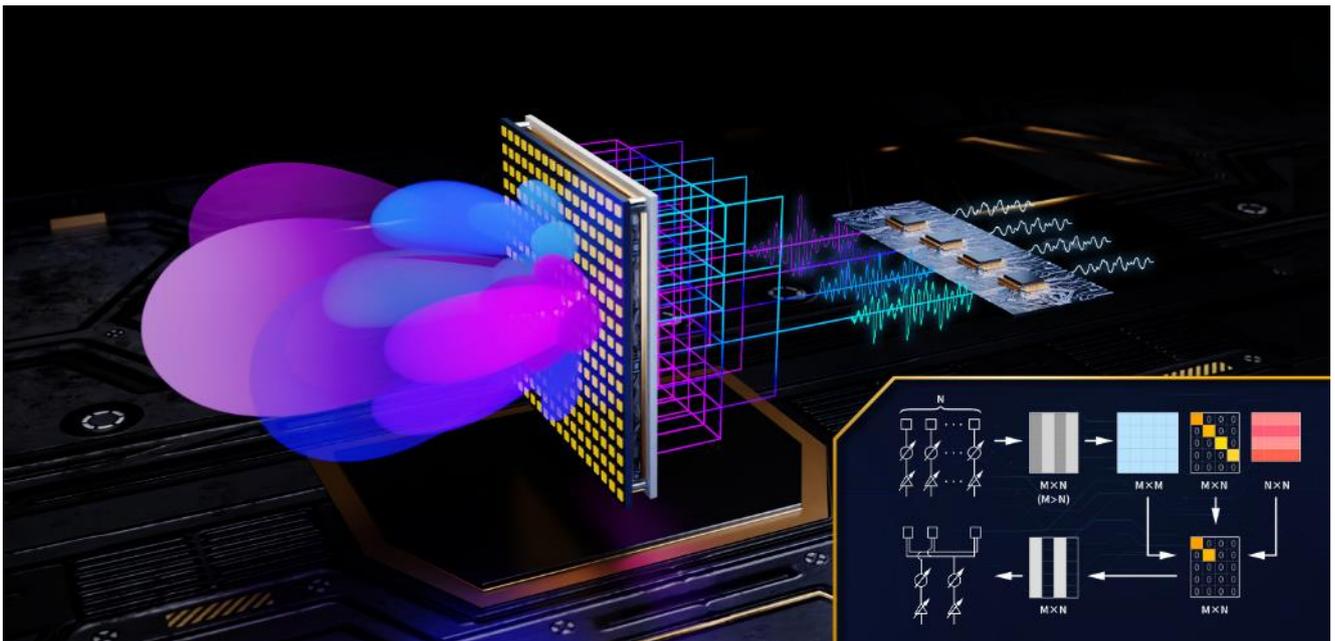
*Figure 1 Dimensionality Reduced Antenna Array Concept*

Beamforming and beam-steering of radio, optical waves, and acoustic wave are critically needed for many applications such as high-capacity radio wireless communication[1], [2], optical wireless communication[3], [4], radar[5], LiDAR (light detection and ranging) [6], arrayed ultrasonic equipment[7], and sonar[8]. Traditionally, the electromagnetic wave is radiated (from a transmitter) or collected (by a receiver) to/from all directions. No matter whether there is an object of interest in a specific direction, the energy is transmitted in/collected from a wide range of directions. By spatially focusing (beamforming) the radio/optical signal to a specified direction, the energy required for realizing a function (e.g. data transmission, data detection, energy transfer) can be efficiently reduced[9]. Moreover, the spatial interference phenomenon between objects radiating/receiving can be minimized by beamforming[10]. It yields, 1) a better spatial de-multiplexing of parallel data streams, enabling a higher data capacity [9], [11]–[13]; 2) a better determination of angular reflection reducing multipath effects which may deteriorate system performance[14]. Using the same total transmit power from multiple antennas, the beamforming can provide a stronger received signal since the signals from different transmitter antennas are coherently summed up. To lead the focused wave to a desired direction, a beam steering function is thus required together with beamforming. The most popular and powerful scheme to enable beam-steering is the well-known phased array antenna (PAA), which is an array of antenna elements that are driven by signals with well-determined phase relations between those elements[15]. Here an antenna refers to a radiator of all kinds of waves which can be regarded as propagating phase fronts, hence including electromagnetic waves (so light waves as well) and acoustic waves.

Traditionally, a 1-by-N phased array antenna requires N-1 phase shifters or other active control units. Many schemes are proposed to reduce the number of phase shifters, but other kinds of active control units should be introduced as the alternative[16]–[19]. Table. 1 shows the state-of-art regarding the number reduction of phase shifters in a PAA. The schemes used in Table. 1 are usually based on decomposed subarrays[20]–[25] and vector synthesis[26]–[29]. These ideas do not change the key challenges of the large number of PAA.

In this paper, we propose a new concept of cascaded angle offset phased array antenna (CAO-PAA) and dimensionality reduced CAO-PAA(DRCAO-PAA). Our concept treats the phase shift steering in all directions as a coefficient matrix. Reducing the number of phase shifters is then translated into reducing the dimensionality of the coefficient matrix. The coefficient matrix of a traditional arrangement of phase shifters is full rank. In other words, dimensionally complete, which guarantees the steering in all directions. However, such a matrix is usually sparse, and its dimensionality can be reduced in many cases. This new insight brings a completely new solution to reducing active components of large-size PAA. As demonstrated in this paper, the 127 phase shifters of a 1-by-128 array are replaced by 16 active control units with similar or lower complexity. Moreover, empowered by the particle swarm optimization（PSO）and Transformer, we introduced the excellent search of the linear independent vector group(LIVG) for a given coefficient matrix. This paves the way for practical application. The numerical and experimental results verify the whole concept of DRCAO-PAA.

*Table 1 State-of-art to reduce phase shifters in PAA*

|  | Optimization | Number of element | Number of phase shifters | Reduction of phase shifters(%) | Beam steering range(°) | Peak side lobe level（dB） |
|---|---|---|---|---|---|---|
| [26] | No | 8 | 2 | 75 | 25 | -10 |
| [27] | No | 5 | 2 | 60 | 16 | -15 |
| [22] | No | 30 | 12 | 60 | 48 | -15 |
| [28] | Yes | 8 | 2 | 75 | 37 | -9 |
| [23] | Yes | 28 | 14 | 50 | 48 | -15 |
| [24] | Yes | 16X16 | 64 | 75 | 40 | 10~12 |
| [25] | Yes | 7 | 3 | 57.1 | 50 | -20 |

**Review on Reduce Active Controller in PAA**

The current state-of-art reduced active controller in PAA is divided into two directions: 1) vector synthesis based reduced active controller, and 2) sub-array compression, using a separate phase shifter for each sub-array.

The earliest vector synthesis phased array was proposed by Amir Mortazawi in 2010[26]. Although the amount of the phase shifter is reduced, each array element is still followed by a tuneable amplifier control. In this study, 8 antenna array elements can complete a 25° scan with 2 phase shifters and 8 tuneable amplifiers. Based on this theory, Amir led his group to conducted several studies[29], [30]. The latest research results of their group have successfully controlled the 8 antenna array elements to complete a $\pm 90°$ scan in $Ku$ band using only 4 adjustable phase shifters[29].

Subarray compression was first proposed in 1974[20]. After that, many scientists started to study the optimal subarray division methods, such as random subarray size [22], interwoven network for subarray[23], recursive algorithm for subarray division[24]. So far the most concise phase shifter networks are introduced in [24], which use 64 phase shifters and 64 adjustable amplifiers, and control the 16X16 array to complete a ±20° beam scan.

The above research has made a remarkable contribution to the reduction of active controllers. However, it still floats on the surface of experiments and lacks a unified theoretical framework and a unified mathematical explanation of the effect of active controller reduction on the system.

**Cascaded angle offset phased array antenna (CAO-PAA)**

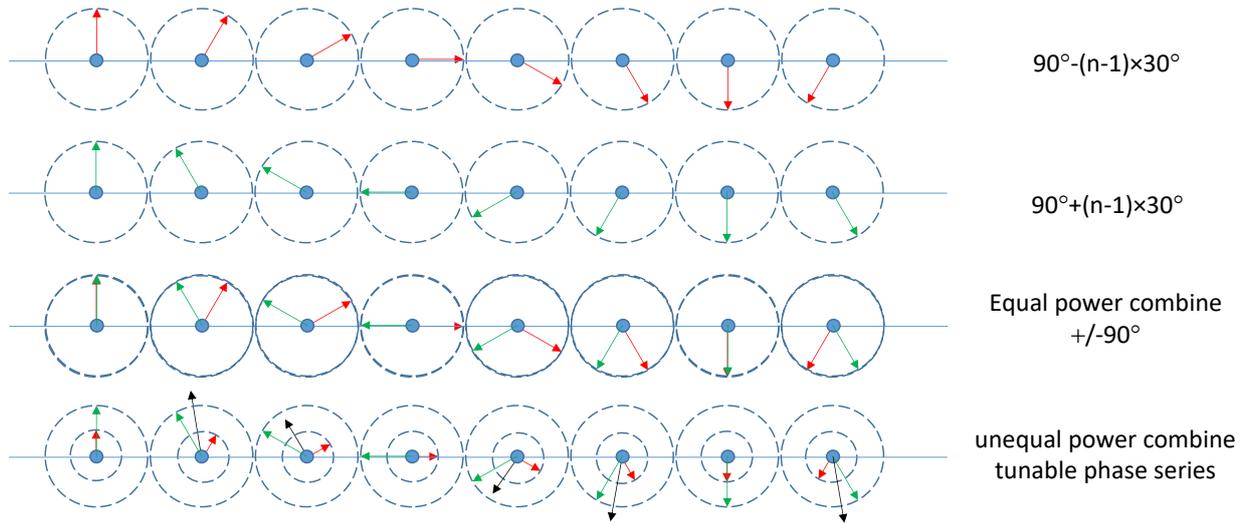

*Figure 2 The basic concept of cascaded angle offset phased array. The green and red arrows represent two cascaded angle offset groups. The black arrows represent the combination of the weighted cascaded angle offset groups.*

Figure 2 illustrates the basic idea of a cascaded angle offset phased array in a intuitive way. Each dashed circle denotes the initial phase of continuous waves at the same carrier frequency. The angles of the arrows denote the value of phases (angle offsets). In the first row, a cascaded angle offset group (CAO-G) is generated with the phase offset of the $nth$ phase shifter denoted as:

$$90° - (n-1) \times 30° \qquad (1)$$

Similarly, the second row denotes another CAO-G generated with:

$$90° + (n-1) \times 30° \qquad (2)$$

The continuous waves with the two CAO-G are then combined by summing them with equal/unequal power ratios as shown in the third/fourth rows. The resulting combined signals are represented by the black arrows. The combined continuous waves are then launched to a linear array of antennas. As shown in the third row of Fig. 2, the power ratio is 1:1, which yields that the generated waves only have two angle offsets, namely $\pm 90°$. As shown in the fourth row, when the power ratio is changed, the combined phases will be tuned quasi-linearly, which fits a linear phased antenna array.

To optimize or replace the current phased array system, it is necessary to explore the theoretical framework of CAO-PA to stretch the innate advantages of the synthesis. Matrix expression and operation provide appropriate solutions.

In this way, the beam directions of a phased array antenna with element spacing $d$ can be determined by a $1 \times N$ linear phase shift group. For a beam direction at $\theta_m$, the corresponding phase shifts at the $n$-th elementary antenna of a PAA can be expressed as:

$$\varphi_n = (n-1)\left(\frac{2\pi}{\lambda}\right) \times d \times sin\theta_m = (n-1) \times \Delta\varphi_m \tag{3}$$

For the phase shifts of all elementary antennas, the phase shifts distribution can be denoted in a vector as shown below:

$$\Phi_{N \times 1} = \begin{bmatrix} 0 \\ \Delta\varphi_m \\ 2\Delta\varphi_m \\ 3\Delta\varphi_m \\ \vdots \\ (n-1) \times \Delta\varphi_m \end{bmatrix} \tag{4}$$

It can also be written as an $N \times M$ matrix as below.

$$\Phi_{N \times M} = \begin{bmatrix} 0 & 0 & \cdots & 0 \\ \Delta\varphi_1 & \Delta\varphi_2 & \cdots & \Delta\varphi_M \\ 2\Delta\varphi_1 & 2\Delta\varphi_2 & \cdots & 2\Delta\varphi_M \\ 3\Delta\varphi_1 & 3\Delta\varphi_2 & \cdots & 3\Delta\varphi_M \\ \vdots & \vdots & \ddots & \vdots \\ (n-1)\Delta\varphi_1 & (n-1)\Delta\varphi_2 & \cdots & (n-1)\Delta\varphi_M \end{bmatrix} \tag{5}$$

Each vector $\Phi_{N \times 1}^M$ can be decomposed into a linear combination of $R$ vectors ($\Phi_{N \times R}$) that can be referred to as the basis of $\Phi_{N \times M}$ where $R$ is the rank of $\Phi_{N \times M}$. Usually, the number of scanning directions $M$ is much larger than the number of elementary antennas $N$. Therefore, $R$ is no larger than $N$. Particularly, when $R$ is equal to $N$, one special basis is the identity matrix. It is the case of a traditional phased array, where the phase and amplitude of the radiated signal at each antenna are independently controlled. For the rest, the number of active controllers can be reduced with the basis $\Phi_{N \times R}$. Moreover, regardless of some weights of the whole basis, the whole system can be further simplified with a new fungible basis $\Phi_{N \times \Lambda}$, where $\Lambda$ is a little bit less than $R$. In this way, a reasonable fungible basis will give rise to the performance penalty which can be reduced accuracy of beam directions or reduced antenna gain in an acceptable range.

**Dimensionality Reduced CAO-PAA**

To achieve the goal of stretching the innate advantages of CAO-PAA, a matrix compression method is applied to simplify the basis of $\Phi_{N \times M}$, avoiding the redundancy of active controllers. Assume that A is an $M \times N$ matrix indicating the input signals of the array elements, as in (6):

$$A = \begin{bmatrix} A_{1,1}e^{j\varphi_{1,1}} & \cdots & A_{1,n}e^{j\varphi_{1,n}} \\ \vdots & \ddots & \vdots \\ A_{m,1}e^{j\varphi_{m,1}} & \cdots & A_{m,n}e^{j\varphi_{m,n}} \end{bmatrix}, \tag{6}$$

where $A_{i,j}$ and $\varphi_{i,j}$ are the amplitudes and phases of each antenna, respectively. *N* is the number of antennas in a PAA, and $M$ is the number of different beam directions. Due to the relations, each row vector denotes one combination of amplitude and phase in an $N$-element antenna array, which indicates one unique direction. With the singular value deposition (SVD)[31], the matrix A can be expressed as

$$A = USV^*, \qquad (7)$$

where U is a $M \times M$ unitary matrix, S is a $M \times N$ rectangular diagonal matrix with non-negative real numbers on the diagonal, and $V$ is a $N \times N$ unitary matrix. The number of non-zero singular values is equal to the rank of $A$. Because of the non-uniqueness of SVD, it is possible to choose the decomposition with the singular values in descending order. The singular values are always ordered from largest to smallest. Once a singular value is found to be less than an arbitrarily small number $\varepsilon$, then the remaining singular values will be replaced by 0 to generate the reasonable fungible basis that is discussed above. The rank of the new diagonal matrix $\Lambda$ equals the number of non-zero terms in the matrix. The assessment of the singular value can be adjusted by the accuracy of the system. In this way, a compressed matrix $B$ can be expressed with proper rank as

$$B = U\Lambda V^*. \qquad (8)$$

With the desired rank, the difference between $A$ and $B$ can be limited to a small range, which minimizes the performance penalty.

After matrix compression, the linear independent vector group (LIVG), which can be determined by matrix $B$, is used to describe the new amplitude and phase distributions. Without loss of generality, assume that the rank of $B$ is equal to $R$. Due to the features of the matrix, $p$ bases can be chosen to express the rest $M-R$ row vectors in matrix $B$. Then all directions from *M* can be expressed in a linear superposition of the vectors in the LIVG matrix.

For any direction in the matrix $B$, there is always a linear vector $K$, which denotes the linear combination of LIVG. Assume that $K$ is a column vector, and the relation of the linear combination is given by:

$$CK = P', \qquad (9)$$

where $C$ is the LIVG matrix with full a rank, and $P'$ is the selected row vector in matrix $B$. Due to the full rank of the augmented matrix in Eq.(11), $K$ has a solution from the system of linear equations. Therefore, based on one LIVG, any direction corresponds to one unique K. Furthermore, all the elements in $K$ are in form of $A_m e^{j\varphi}$, where $A_m$ is the amplitude, and $\varphi$ is the phase. The practical function of $K$ can be realized by $R$ amplifiers and $R$ phase shifters. In this way, the complexity of the $M \times N$ matrix in the whole system can be decreased by using a LIVG and $R$ active controllers.

**Dimensionality Reduced CAO-PAA in Beam Steering**
Phase shifts and amplitude weights are applied to a 1-D linear evenly spaced phased array antenna array in order to evaluate the beam steering performance utilizing the proposed DRCAO-PAA concept. At angle $m$, weights with array number $N$ reconstructed by SVD compression can be expressed

as:

$$P'_m = CK = k_1 e^{-j\kappa_1}\vec{c}_1 + k_2 e^{-j\kappa_2}\vec{c}_2 + \cdots k_r e^{-j\kappa_r}\vec{c}_r, \tag{10}$$

where $k_r$ denotes the *rth* magnitude, $\kappa_r$ denotes the *rth* phase, $\vec{c}_r$ denotes the *rth* row vector in the composed LIVG matrix, which can be expressed as

$$\vec{c}_r = \begin{bmatrix} C_{r,1} e^{(-j\phi_{c(r,1)})} \\ C_{r,2} e^{(-j\phi_{c(r,1)})} \\ \vdots \\ C_{r,n} e^{(-j\phi_{c(r,n)})} \end{bmatrix}, \tag{11}$$

where $C_{r,n}$ denotes the amplitude of the *rth* vector at *nth* element, $\phi_{c(r,n)}$ denotes the phase shifts of the *rth* vector at *nth* antenna.

Eq.(10) can be rewritten as

$$P_m' = \begin{bmatrix} C_{1,1} k_1 e^{-j\kappa_1 \phi_{c(1,1)}} + C_{2,2} k_2 e^{-j\kappa_2 \phi_{c(2,1)}} + \cdots + C_{r,1} k_r e^{-j\kappa_r \phi_{c(r,1)}} \\ C_{1,2} k_1 e^{-j\kappa_1 \phi_{c(1,2)}} + C_{2,2} k_2 e^{-j\kappa_2 \phi_{c(2,2)}} + \cdots + C_{r,2} k_r e^{-j\kappa_r \phi_{c(r,2)}} \\ \vdots \\ C_{1,n} k_1 e^{-j\kappa_1 \phi_{c(1,n)}} + C_{2,n} k_2 e^{-j\kappa_2 \phi_{c(2,n)}} + \cdots + C_{r,n} k_r e^{-j\kappa_r \phi_{c(r,n)}} \end{bmatrix}. \tag{12}$$

The direction vector of a one-dimensional phased array, denoted as

$$d(\theta) = [1, e^{jkd\cos(\theta)}, e^{j2kd\cos(\theta)}, \ldots e^{j(N-1)kd\cos(\theta)}], \tag{13}$$

$k = 2\pi/\lambda$, $\lambda$ is the RF signal wavelength.

Based on beamforming theory[32], The far field pattern (FFP) of such an array could be expressed as:

$$AF(\theta) = |d(\theta) P'_m| = |\sum_{n=1}^{N-1} \sum_{r=1}^{R} C_{r,n} k_r \exp(-j(\kappa_r \phi_{c(r,n)} + nkd\cos(\theta))|. \tag{14}$$

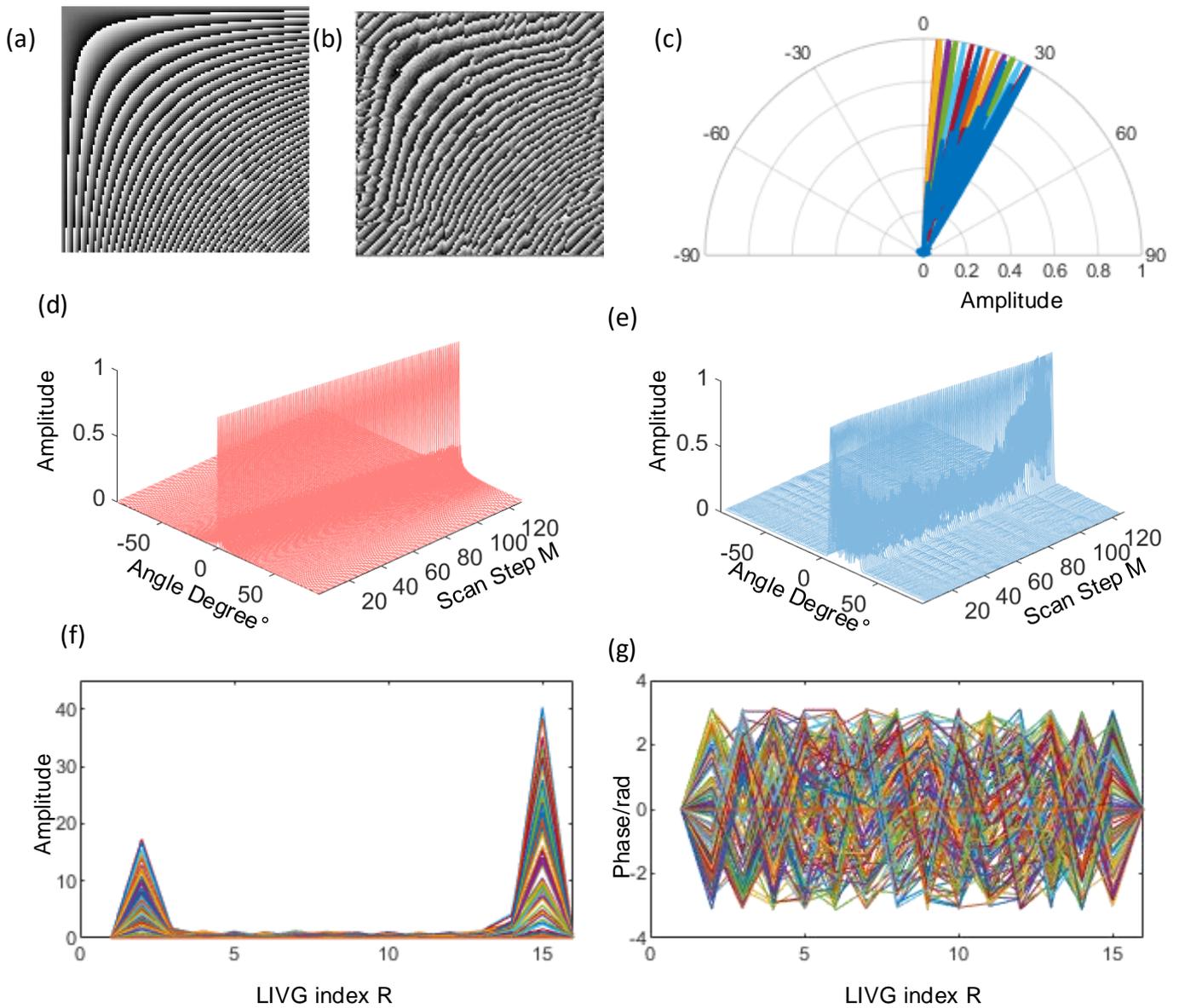

*Figure 3 Simulation results for 30° scan with N=128 and the LIVG with 16 vectors. (a)Gray scale map of the phase shifts in array N=128 elements for M=128 angles; (b) Gray scale map of the phase shifts in array N=128 for M=128 angles after SVD compression; (c) Normalized far-field pattern for LIVG consists of equally spaced vectors; (d) Normalized Ideal far-field pattern from 0° to 30°; (e) Reconstructed matrix normalized far-field pattern from 0° to 30°; (f) Magnitude of K factors for each LIVG vectors; (g) Phase of K factors for each LIVG vectors.*

The simulation results in Fig. 3(a)–(g) show that the beam steering from 0 to 30° can be realized with just 16 active controllers in controlling a $1\times 128$ array. The phase shifts of each element of the one-dimensional 128 array $N=128$ for 128 different scan angles $\theta_m$ ($0<\theta_m<30°$) are shown in Fig.3(a). A grayscale map is used to depict it; for instance, a grayscale of 0 (black) indicates -180° phase shifts, whereas a grayscale of 1 (white) indicates +180°. After SVD compression, the initial 128 eigenvalues are reduced to 16 eigenvalues, and the phase shifts matrix is reconstructed as shown in Fig.3(b).

To illustrate the process of dimensionality reduced antenna array in beam steering in more details, the beam steering simulation was performed with 16 vectors of the reconstructed matrix selected equally spaced. These 16 vectors were used to compute and synthesize a far-field pattern, which was shown in Fig. 3(c). Comparing the normalized far-field pattern calculated with (Fig. 3(e)) and without (Fig. 3(d)) eigenvalue compression matrix, the mean absolute error of the beam pointing is only within 1.3°. This indicates that there is very little beam pointing loss due to the distortion appeared in Fig. 3(b).

The simulation analysis shows that it is possible to choose vectors at equal intervals to create LIVG. However, different vector selection functions will affect the factors of $K$. Fig. 3(f)-(g) illustrates the amplitude and phase of $K$ while scanning 128 various angles, which indicate the magnification required by the amplifier and the range of the phase shifter to be adjusted. Especially the amplifier multiplier, the maximum value exceeds 40. The required phase shift is evenly distributed between [-180°, 180°]. Therefore, LIVG algorithms are proposed to optimize the LIVG building.

**LIVG Building in DRCAO-PAA**

From theoretical analysis there are infinite sets of solutions LIVG matrix, however only part of them can be used in applications to avoid obtaining improper K with an extremely high amplification factor. Considering the cost and benefit of an array in applications, the amplifier gain is constrained, and the phase shifts precision is also limited. Thus, the establishment of LIVG affects the performance of the DRCAO-PAA beam steering.

The most straightforward and effective way is to directly choose $R$ linearly uncorrelated vectors in the $M \times N$ matrix to build the LIVG such as 16 equally spaced vectors displayed in Fig.3(c). In order to find out the proper LIVG, particle swarm optimization (PSO) is introduced in our algorithm[33]. An optimization function is provided to PSO to minimize the performance penalty in beam steering:

$$\min imize(\sum_{m=1}^{M}(\text{abs}(Amp_{idear,m} - Amp_{svdcompr,m})) + (abs(Angle_{ldear,m} - Angle_{svdcompr,m})))/2M \quad (15)$$

$$Subject \quad to \quad \max(abs(K)) < 20$$

At $mth$ scan step, $Amp_{idear,m}$, $Angle_{ldear,m}$ represent the main lobe magnitude and pointing angle of the ideal far-field pattern; $Amp_{svdcompr,m}$, $Angle_{ldear,m}$ represent the main lobe magnitude and pointing angle of the LIVG reconstructed far-field pattern. The goal of the optimization is to minimize the average difference between the ideal far-field pattern and the LIVG reconstructed far-field pattern in terms of main lobe beam pointing and magnitude. The constraint is that the amplifier gain needs to be less than 20(~13dB). This approach minimizes errors in beam pointing and main lobe amplitude when performance penalties are inevitable, to enhance DRCAO-PAA 's beam steering performance.

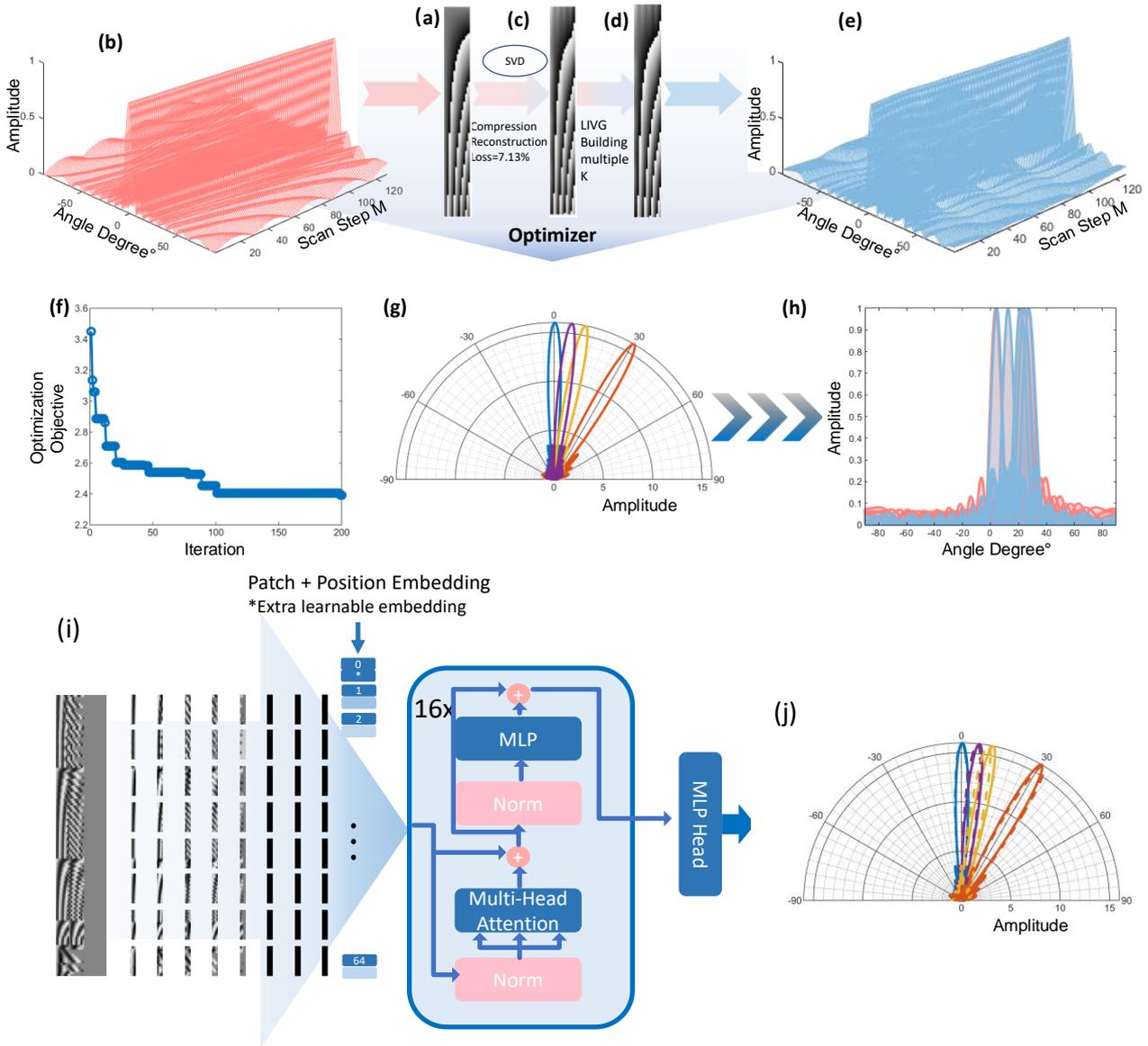

*Figure 4 Schematic diagram LIVG Building and Optimization: PSO case in scan angle 0° to 30°, each step 30°/128. Deep Learning Case input data set consist of scan angle=m(m=15,20,···,85); element=n(n=4,5,6,···,16), rank=n-1;output data set was the optimized LIVG. (a)M=128, N=16 Grey-scale map represents phase shifts [-180°,180°]at element n, degree m; (b) Normalized Ideal Far-field pattern form 0° to 30°; (c) Grey-scale map after SVD compressed. (d) Grey-scale map reconstructed by LIVG. (e) Reconstructed weight matrix far-field pattern from 0° to 30°. (f) Iterative convergence of PSO algorithm. (g) The optimally constructed LIVG, where each row vector of the LIVG represents a specific beam pointing. (h) Normalized Ideal far-field pattern and optimal LIVG-built normalized far-field pattern. (i) Deep Learning Data Set building; (j)Topology of Transformer Deep Learning Network Structure; (k) Far-field pattern of the LIVG matrix generated by the deep learning model (dashed line) and PSO algorithm (solid line).*

As seen in Fig. 4(a) -(h), the process of obtaining the optimal LIVG is demonstrated by using the PSO algorithm. The simulation is based on a 16 elements ($N=16$) line phased array with a maximum scan angle of 30°, and the number of scanning steps is 128 ($M = 128$). Fig. 4(a) displays the weights for the 16 array elements as they are scanned at various angles. An ideal linear antenna array could synthesize beams at a specific angle based on the weights of each row, as shown in Fig. 4(b). Considering the performance penalty, an appropriate degree of compression should be set after SVD. In this case, the first four eigenvalues make up 92.87% of the weight of the total eigenvalues.

Therefore, in this case four vectors are selected to construct LIVG ($rank = 4$). After recreating the matrix by zeroing the remaining eigenvalues, as illustrated in Fig.4(c), the distortion of the weights is almost negligible. The weight matrix as in Fig. 4(d) is reconstructed again after applying the constraints of Eq. (15) to the coefficients $K$. The distortion of the weight matrix gradually becomes apparent. Such distortion could potentially affect the beam pointing.

The optimizer's objective is to create a better collection of LIVGs such that the difference between the ideal far-field pattern [Fig.4(b)] and the reconstruction's far-field pattern [Fig.4 (e)] is sufficiently minimal to improve beam pointing accuracy. The algorithm is expected to converge the optimizer function value at 2.39 after 200 iterations as shown in Fig. 4(f). As shown in Fig. 4(g) the optimized LIVG was built, and the far-field pattern formed by the restructure weight matrix is compared to the far-field pattern of the ideal antenna in Fig. 4(h). The reconstructed far-field pattern is nearly identical to the ideal far-field pattern during the 30° scan in terms of beam direction, which implies that the active devices of the array system can be cut down to 25% by regulating the phased array of 16 elements using only 4 amplifiers and phase shifters and loss the beam pointing accuracy less than 1.73°.

To eliminate the redundancy of the repetitive computation and design a general LIVG building model, a deep learning model was designed. A wider field-of-view model is required due to our data consists of a collection of numeric values. Transformer[34] has the qualities we need, such as a relatively low induction bias and a correlation of all feature values, compared to Convolutional Neural Networks(CNN)[35] and Recurrent Neural Networks (RNN)[36].

To generate training data for this model, as depicted in Fig. 4(i), a more comprehensive dataset was constructed by traversing scanning angles from 5° to 85° with 5° for each step, element numbers from 4 to 16 with 1 for each step, and LIVG with rank $r = 2$ to $N-1$ at intervals. For each piece of data, the input is a $128 \times 32$ matrix that includes the 16 array components' complex weights (both real and imaginary) as well as the number of scan steps of M=128 (If the number of arrays N<16, the remaining values are zero. The output is the optimized LIVG for each input matrix.

The topology of the transformer is very simple, as shown in Figure 4(j). First, the weight matrix is expanded by the visual transformer into 64 blocks of size $64 \times 16$. In contrast to the standard visual transformer[37], patches are one-dimensional data in DRCAO-PAA, so that the Transformer, which is commonly used to process languages, can each patch be considered as a token (like a word) that can be inserted into the Transformer.

As the model performance is admirable in the test dataset with the 89.68% cosine similarity and the 0.0748 MAE, the Transformer-based model can quickly generate an optimum LIVG, when given the

DRCAO-PAA design parameters (such as the number of array elements, maximum scan angle, etc.). Fig.4(k) depicts the far-field pattern of the LIVG matrix generated by the deep learning model (dashed line) and the far-field pattern of the optimal LIVG matrix solved by the PSO algorithm (solid line). The deep neural network fits the PSO algorithm well. It will simplify the design process and speed up the DRCAO-PAA calculations.

**System demonstration**

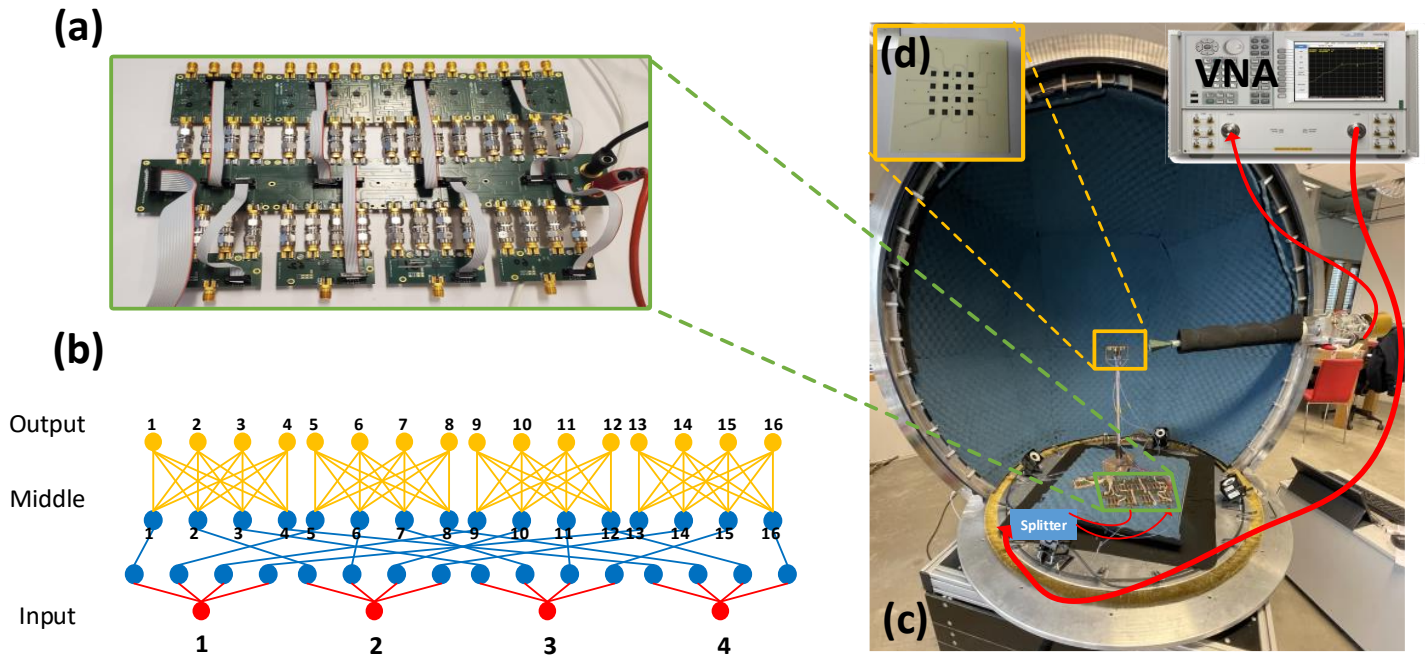

Figure 5 a) DRCAO-PAA verified board prototype b) DRCAO-PAA verified board Schematic c) Far-field Pattern Measurement in Spherical mm-Wave Anechoic Chamber d) 28GHz mm-wave phased array

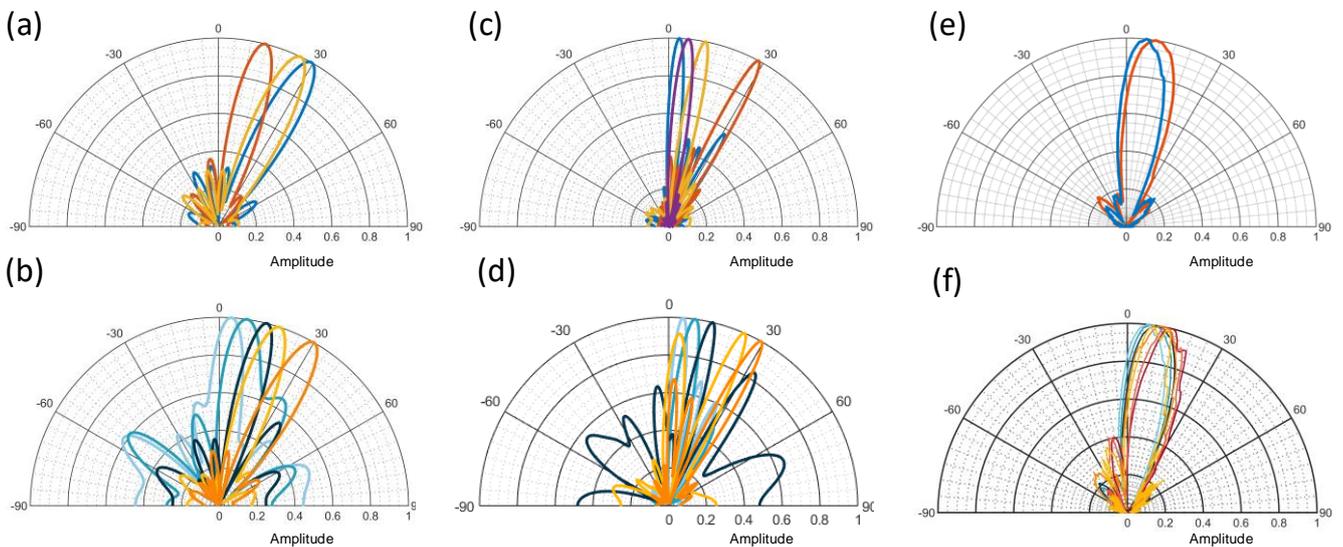

Figure 6 Experimental results **a)** Calculate and synthesize far-field pattern for 3 vectors in LIVG with r=3 **b)** Control LIVG with r=3 to complete 0~30° beam steering **c)** Calculate and synthesize far-field pattern for 4 vectors in LIVG with r=4 **d)** Control LIVG with r=4 to complete 0~30° beam steering **e)** Measured far-field pattern for 2 vectors in LIVG with r=2 **f)** Control LIVG with r=2 to complete 6°~14°beam steering

As shown in Fig. 5(a), a mm-wave phase shifter PCB module is used to verify the DRCAO-PAA theory. Fig.5(b) shows the conceptional block diagram of the phase shifter module. Three layers—the input layer, middle connection layer, and output layer—combine to form the entire module design. In this experiment, a maximum of 4 inputs and 16 outputs were accomplished. By having low-cost and commercially available silicon beamformer chips on the input and output layer PCBs, each path between input and output ports can adjust its amplitude and phase independently with high resolution.

For mapping DRCAO-PAA theory, each yellow tap represents a fixed tap, to implement the weights of each vector in the LIVG; red features a tuneable tap with adjustable amplitude and phase (linear Vector K). As an illustration, consider a LIVG built with rank=2 and N=4. The LIVG weights can be set by adjusting the output chip's control of the amplitude and phase of the link between *Middle 1-Output 1,2,3,4* and *Middle 2-Output 1,2,3,4*. Beam steering is achieved by adjusting the amplitude and phase of the link between *Input 1-Middle 1* and *Input 2-Middle 2* with variable K.

A LIVG with rank=3 N=8(referred to as r=3) which reduces 62.5% component, and a LIVG with rank=4 N=16(referred to as r=4) which reduces 75% component, were both utilized to evaluate the DRCAO-PAA. After finishing the fixed tap equalization via the SPI serial interface. To achieve the 0–30° beam steering, r=3 or r=4 taps are employed to control 3 or 4 vectors of LIVG. The S21 parameters were measured between each input port and output port on the DRCAO-PAA verification board with a vector network analyzer (VNA Agilent Technologies PNA Network Analyzer E8361C).

The normalized far-field pattern of the beam was synthesized by the calculation of measured $S_{21}$ parameters. Fig.6 (a) shows the normalized far-field pattern computed and synthesized from each of the three vectors in the LIVG with rank=3, pointing at 14.4°, 25.92°, and 29.52°, respectively. Five sets of K-factors were employed to control the LIVG to achieve beam steering from 0 to 30°. As shown in Fig. 2 (b), the normalized far-field pattern computed and synthesized by 5 sets measured $S_{21}$. The beam steering from 0 to 30° is accomplished based on the DRCAO-PAA theory to regulate these three non-uniformly selected vectors.

For the 4 vectors case, Fig.6(c) shows the normalized far-field pattern computed and synthesized from each of the 4 vectors in the LIVG, pointing at 3.24°, 6.12°, 11.16°, and 28.44°, respectively. With 5 sets of K factors, the beam steering was performed from 0 to 30°. As shown in Fig. 6(d), the normalized far-field pattern computed and synthesized by 5 sets measured S21. The case of 4 vectors is also complete the beam steering from 0 to 30°, and the beam synthesized by 16 antennas is narrower and more accurate pointing.

To further verify the DRCAO-PAA hypothesis, the setupshown in Fig. 5(c) was utilized to evaluate the antenna's far field. The whole measurement was performed in a spherical mm-wave anechoic chamber. As shown in Fig. 6(e)(f), the normalized far-field pattern of the beam was measured using this setup. With rank=2, N=4, the normalized far-field pattern of the two vectors in LIVG, pointing at 6° and 9°, was shown in Fig. 6(e). To achieve the beam steering from 6° to 14°, different K is applied with the help of 2 phase shifters, and 2 variable gain amplifiers (VGA). As shown in Fig. 6(f), six beams are pointing at 6°,7°,8°,9°,10°, and 14°, which demonstrate beam steering from 6° to 14°. This experiment verified the correctness of DRCAO-PAA theory by controlling a 1X4 phased array with two active controllers to complete beam steering.

The experiments demonstrate that, based on DRCAO-PAA theory, any array size of a phased array can be perfectly driven with fewer active controllers to complete beam steering.

**Conclusion**

We propose a novel theory for dimension-reduced antenna arrays. In dimension-reduced cascaded angle offset phased array theory, the Singular value decomposition is introduced to compress the rank of the matrix and compress the base of the scan and weight matrix space to reduce active components. In various application scenarios, various dimensions can be compressed in accordance with various performance penalties. Meanwhile, the particle swarm optimization algorithm and Transformer are used to optimize and accelerate the building of the basis of the scan and weight matrix space.

A series of experiments are used to verify our theory. 3-vectors($rank = 3$ independent vector group) are used to characterize the vector space consisting of 8($N = 8$) antenna array elements weights and 4-vectors($rank = 4$ independent vector group) are used to characterize the vector space consisting of 16($N = 16$) antenna array elements. These experiments are well implemented by testing the weights and calculating the synthetic beam steering for beam scanning from 0 to 30°. Further, two phase shifters and two adjustable amplifiers are employed in a spherical anechoic darkroom to control 4 antenna array elements to complete a continuous beam sweep of 8°.

We experimentally demonstrate the generalizability of our proposed theory. This Dimensionality Reduced Antenna Array theory unifies previous studies on phase-shifter-reduced phased arrays and can be used to analyse all research scenarios that require a reduced active controller. It is a crucial contribution to the design of low-cost, low-complexity, control systems.

**Method**

*Particle swarm algorithm in* DRCAO-PAA

The particle swarm algorithm is a swarm intelligence algorithm designed by simulating the predatory behaviour of a flock of birds. Each particle acts as a potential solution with its unique velocity and position, and each particle has an objective value that is based on its own location which indicated the vectors index of the *M×N* matrix.

First, a set of population particles is produced at random. The particles can alter their own location, which alters the objective value, by keeping track of these two best positions: the population's overall best position gBest and the historical best position of each population particle pBest. The subsequent equation accounts for the particle's position and velocity.

$$\begin{aligned} v^{t+1}_{i,r} &= \omega \times v^t_{i,r} + c_1 \times rand() \times (pBest^t_{i,r} - x^t_{i,r}) \\ &= c_2 \times rand() \times (gBest_r - x^t_{i,r}) \end{aligned} \quad (16)$$

$$x^{t+1}_{i,r} = x^t_{i,r} + v^{t+1}_{i,r} \quad (17)$$

where i = 1,2,......,N, N is the seed size, r is the dimensionality which is the rank of the LIVG in this algorithm, v is the particle velocity, x is the particle position which is the index in the *M×N* matrix, v is the particle velocity, t is the current iteration number, and c1, c2 are learning factors, and rand() is a uniformly distributed function between [0,1].

*Transformer for LIVG building*

16 layers of Transformer Encode were used to build our model which proposed neither a CNN nor an LSTM, but a mechanism called dot-product Attention, on top of which the model (Transformer) is already significantly better than existing methods.

Query, Key, and Value are the three variables that The Transformer employs in (dot product) attention. In essence, the algorithm multiplies each Key by its corresponding Value to get the Attention Weight of the Query and Key phrases.

$$Attention(Q, K, V) = soft\max(\frac{QK^T}{\sqrt{d_k}})V \tag{18}$$

In our model, Multi-Head Attention is used as well. It uses multiple Attention Heads (in the case of MLP, the "number of hidden layers" is increased), defined as follows:

$$MultiHead(Q, K, V) = Concat(head_1, ..., head_h)W^O \tag{19}$$

where
$$head_i = Attention(QW_i^Q, KW_i^K, VW_i^V) \tag{20}$$

Each Head has its own projection matrices $W_i^Q, W_i^K, W_i^V$, projected features of these matrices are used to create attention.

The Transformer model's training process is split into coarse and fine modes. With the learning rate of 1e-5, the 200 training sessions in coarse mode are carried out using the AdamW optimizer[38]. With the learning rate of 1e-7 and the momentum factor set to 0.1, the 300 training sessions in fine mode are carried out using the SGD optimizer[39]. To prevent overfitting, the mean absolute error (MAE) is used as the loss function of the model. The "cosine similarity," which measures how similar two numerical sequences are, is selected as the metric. The model completes training with a lossy MAE of 0.0562, and a cosine similarity of 91.83%.

*Design of mm-wave phase shifter module*

The mm-wave phase shifter module is operating at 28 GHz, and contains 3 layers (Fig 5a):

-The input layer has 4 inputs and 16 outputs and is made using four 1-to-4 phase shifter PCBs. On each PCB, the input signal is split into 4 paths (1-to-4), where each path has an independent 6 bits 30dB amplitude control and 6 bits 360-degree phase control. This is achieved by using a commercially available beamformer chip, which is set to TX mode in this experiment. The chip operates at 25-30 GHz and has a max output power of +10 dBm at each output. The gain and amplitude control can be done through SPI, and the chip on each PCB has a different hardware address, so that they can be controlled by a single SPI master device.

-The output layer has 16 inputs and 16 outputs and is made using four 4-to-4 PCBs. Each PCB has 4 beamformer chips, which is the same chip used on the input layer PCB. In this case, the beamformer chips are set to RX mode, where each chip can combine 4 inputs signal to 1 output and have independent amplitude and phase control on each path. As a result, by having 4 chips on board in parallel and sharingthe same 4 inputs, each board can operate as a 4-to-4 network.

-The middle layer is built to route and distribute the signals between the input and output layers. It also serves as a master SPI controller and power supply for all the beamformer chips used in the module. The complete mm-wave phase shifter module consumes about 12 W from a single 3.3V power supply. On the board, there are LDOs to regulate the supply voltage and SPI buffer to minimize the noise on the control signals. The SPI can operate up to 50 MHz, which allows the module to synthesize any phase shift in any paths in microseconds.

As each path in the input and output layers has an independent amplitude and phase control, the module provides enough flexibility to synthesize and implement different DRCAO-PAA algorithms, as well as calibrate any system amplitude imbalance between different antenna channels.

**Author contribution**
Z. Cao conceived the idea of DRCAO-PAA, the use of AI to empower the LIVG search process, and led the theoretical analysis and the numerical simulation. S. Xia led the experimental research, designed the AI algorithm, and performed the numerical simulation. M. Zhao contributes to the theoretical analysis, numerical simulation, and design of the antenna array. L. Yang and X. Zhang led and performed the research on the antenna array. L. Yang supervised the research on the antenna array. Q. Ma led and performed the design and realization of the phase shifter board. H. Chung contributed to the design of the phase shifter board. Ad Reniers led the far-field pattern measurement of DRCAO-PAA. A.M.J. Koonen led the whole research and guided the paper writing. All authors contributed to the data discussion and wrote the paper.

**Data availability**
The data that support the plots within this paper and the other finding of this study are available from the corresponding author upon reasonable request.

**Supplement**
2D DRCAO-PAA
In a two-dimensional (2D) array, new phase relation will be introduced to the matrix. In a standard $N \times N$ phased array, in a beam direction $m$, the phase shift $\Delta\varphi_m$ can be expressed as

$$\Delta\varphi_m = (\frac{2\pi}{f}) \times \tau_{m,} = (\frac{2\pi}{f}) \times \frac{d}{c} \times (sin\theta_m \cos\phi_m + sin\theta_m \sin\phi_m)$$
$$= (\frac{2\pi}{\lambda}) \times d \times (sin\theta_m \cos\phi_m + sin\theta_m \sin\phi_m) \quad (21)$$

where d is the distance between array elements, $\theta_m$ and $\phi_m$ are the 2D azimuth angles. Therefore, the phase shifts of the phased array can be expressed as

$$\varphi_{i,j} = (i-1)(\frac{2\pi}{\lambda}) \times d \times sin\theta_m \cos\phi_m + (j-1)(\frac{2\pi}{\lambda}) \times d \times sin\theta_m \sin\phi_m$$
$$= (\frac{2\pi}{\lambda}) \times d \times [(i-1)sin\theta_m \cos\phi_m + (j-1)sin\theta_m \sin\phi_m] \quad (22)$$
$$= (i-1)\Delta\varphi_{m1} + (j-1)\Delta\varphi_{m2}$$

Assume that $\Delta\varphi_{m1} = (\frac{2\pi}{\lambda}) \times d \times sin\theta_m \cos\phi_m$, $\Delta\varphi_{m2} = (\frac{2\pi}{\lambda}) \times d \times sin\theta_m \sin\phi_m$, the 2D distribution can be denoted in a vector as shown below:

$$\Phi_{m,N\times N} = \begin{bmatrix} 0+0 & 0+\Delta\varphi_{m2} & \cdots & 0+(n-1)\Delta\varphi_{m2} \\ \Delta\varphi_{m1}+0 & \Delta\varphi_{m1}+\Delta\varphi_{m2} & \cdots & \Delta\varphi_{m1}+(n-1)\Delta\varphi_{m2} \\ \vdots & \vdots & \ddots & \vdots \\ (n-1)\Delta\varphi_{m1}+0 & (n-1)\Delta\varphi_{m1}+\Delta\varphi_{m2} & \cdots & (n-1)\Delta\varphi_{m1}+(n-1)\Delta\varphi_{m2} \end{bmatrix}. \quad (23)$$

Conveniently, the matrix can be reordered into a $1 \times N^2$ vector,

$$\Phi_{m,N^2 \times 1} = \begin{bmatrix} 0 \\ \vdots \\ 0 + (n-1)\Delta\varphi_{m2} \\ \Delta\varphi_{m1} + 0 \\ \vdots \\ \Delta\varphi_{m1} + (n-1)\Delta\varphi_{m2} \\ \vdots \\ \vdots \\ (n-1)\Delta\varphi_{m1} + 0 \\ \vdots \\ (n-1)\Delta\varphi_{m1} + (n-1)\Delta\varphi_{m2} \end{bmatrix}. \tag{24}$$

Like the 1D case, we can also combine $m$ vectors into a group of linear vectors as below,

$$\Phi_{N^2 \times M} = \begin{bmatrix} \Phi^1_{N^2 \times 1} & \Phi^2_{N^2 \times 1} & \ldots & \Phi^M_{N^2 \times 1} \end{bmatrix}. \tag{25}$$

Compared with Eq.(22), the new vectors can be written as below.

$$\Phi_{N^2 \times M} = \begin{bmatrix} 0 & 0 & & 0 \\ \vdots & \vdots & & \vdots \\ 0+(n-1)\Delta\varphi_{12} & 0+(n-1)\Delta\varphi_{22} & & 0+(n-1)\Delta\varphi_{m2} \\ \Delta\varphi_{11}+0 & \Delta\varphi_{21}+0 & & \Delta\varphi_{m1}+0 \\ \vdots & \vdots & & \vdots \\ (n-1)\Delta\varphi_{11}+(n-1)\Delta\varphi_{12} & (n-1)\Delta\varphi_{21}+(n-1)\Delta\varphi_{22} & \ldots & (n-1)\Delta\varphi_{m1}+(n-1)\Delta\varphi_{m2} \\ \vdots & \vdots & & \vdots \\ \vdots & \vdots & & \vdots \\ (n-1)\Delta\varphi_{11}+0 & (n-1)\Delta\varphi_{21}+0 & & (n-1)\Delta\varphi_{m1}+0 \\ \vdots & \vdots & & \vdots \\ (n-1)\Delta\varphi_{11}+(n-1)\Delta\varphi_{12} & (n-1)\Delta\varphi_{21}+(n-1)\Delta\varphi_{22} & & (n-1)\Delta\varphi_{m1}+(n-1)\Delta\varphi_{m2} \end{bmatrix} \tag{26}$$

Therefore, similar results can be obtained in a 2D array to reduce the number of active controllers. The 2D array can be considered a group of 1D arrays. As described above, 1D beam steering can be realized by operating a 1D array with the beam angle $\theta$. And 2D array will introduce the new control parameter in the orthogonal direction, where the amplitude and phase shifts among the groups will determine the specific beam angle $\phi$.

**Simulation Result of 2D** DRCAO-PAA
The simulation results of 2D DRCAO-PAA are shown in Fig. S1. which is based on a 4X4 2D phased array with LIVG with rank=4. The beams synthesized by the compressed(blue) weights and the beams synthesized by the pre-compressed(red) weights almost completely overlap during the pitch and direction angles scanned from 0~30°, respectively.

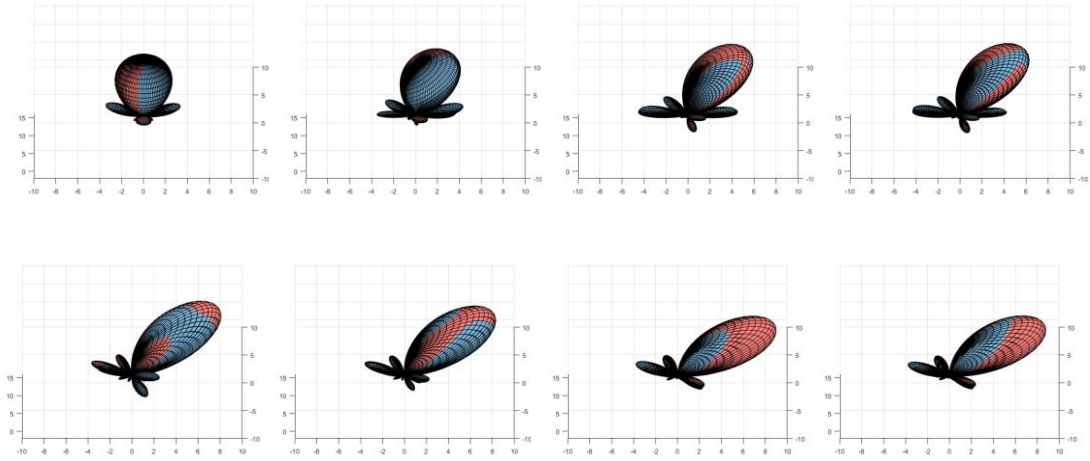

*Figure S1 Steering process of the beam from pitch angle 0 to 30° (first row) and azimuth angle 0 to 30° (second row)*